\documentclass[preprint,12pt]{elsarticle}
\usepackage{amssymb}
\usepackage{graphicx}
\usepackage{dcolumn}
\usepackage{amssymb,amsmath}
\usepackage{natbib}
\usepackage{subfigure}
\usepackage{placeins}
\usepackage{color}

\journal{Renewable Energy}

\begin{document}

\begin{frontmatter}

\title{A concurrent precursor inflow method for Large Eddy Simulations and applications to  finite length wind farms}
\author{Richard J. A. M. Stevens$^{1,2}$, Jason Graham$^{1}$ and Charles Meneveau$^{1}$}
\address{

$^1$Department of Mechanical Engineering, Johns Hopkins University, Baltimore, Maryland 21218, USA\\
$^2$Department of Physics, Mesa+ Institute, and J.\ M.\ Burgers Centre for Fluid Dynamics, University of Twente, 7500 AE Enschede, The Netherlands}

\date{\today}

\begin{abstract}
In order to enable simulations of developing wind turbine array boundary layers with highly realistic inflow conditions a concurrent precursor method for Large Eddy Simulations is proposed. In this method we consider two domains simultaneously, i.e.\ in one domain a turbulent Atmospheric Boundary Layer (ABL) without wind turbines is simulated in order to generate the turbulent inflow conditions for a second domain in which the wind turbines are placed. The benefit of this approach is that a) it avoids the need for large databases in which the turbulent inflow conditions are stored and the correspondingly slow I/O operations and b) we are sure that the simulations are not negatively affected by statically swept fixed inflow fields or synthetic fields lacking the proper ABL coherent structures. Sample applications are presented, in which, in agreement with field data a strong decrease of the power output of downstream wind-turbines with respect to the first row of wind-turbines is observed for perfectly aligned inflow.
\end{abstract}


\end{frontmatter}

\section{Introduction} \label{section1}
{\color{black} Due to the large separation of scales, wind farm studies have often relied on computational efficient models. Examples include the Jensen or PARK model, in which the wakes are represented by an expanding region with a velocity deficit \cite{jen84,kat86}; parabolized forms of the Reynolds-Averaged Navier Stokes (RANS) equations such as the Ainslie model \cite{ain88} and UPMPARK, which uses a $k-\epsilon$ closure, models which are based on a parametrization of the internal boundary layer growth coupled with some eddy viscosity model (e.g.\ the Deep-Array Wake Model \cite{bro12} and the Large Array Wind Farm model \cite{has09}); and linearized CFD models such as FUGA \cite{ott11}, Windmodeller \cite{bea12}, and Ellipsys \cite{iva08}. For reviews of these different methods we refer to \cite{ver03,cre99,bar09b,ret09,san11,cab11,bea12}. Although these models are attractive for their quick runtime, they cannot predict a wealth of important unsteady phenomena in the way Large Eddy Simulations (LES) can. Starting with LES of simple wind turbines using the actuator disk model [Jimenez et al.\ \cite{jim07}], recently such LES of the interaction between wind {\color{black}farms} and the turbulent Atmospheric Boundary Layer (ABL) have become available. Meyers and Meneveau \cite{mey10}, Calaf et al.\ \cite{cal10,cal11}, and Yang et al.\ \cite{yan12} performed LES in a horizontally periodic domain in order to study infinitely long wind-farms.} {\color{black} Ivanell \cite{iva10} and Churchfield et al.\ \cite{chu12,chu12b,lee12} performed LES of the Horns Rev and Lillgrund wind farm, respectively.  Wu and Port\'e-Agel \cite{wu11,wu13} did LES of finite length wind-farms to study the downstream development for aligned and staggered configurations.} In LES, high-fidelity numerics such as spectral methods in horizontal planes is often preferred, with the limitation to domains with periodic boundary conditions. Traditionally, when one uses inflow data from a precursor simulation to model the turbulent inflow, it is required that a precursor simulation is completed before the actual simulation in the domain of interest can be started. This method has been applied in some earlier LES studies of finite length wind farms {\color{black}\cite{chu12b,lee12,wu13}.} During the precursor simulation data are sampled and written to file so that they can be used as turbulent inflow condition later. In the actual simulation the inflow data are read from the database. This approach is conceptually simple, but it has several practical drawbacks. First, the precursor simulation needs to be completed before the actual simulation can be started. Another computational drawback is that a database with inflow conditions requires extensive disk space, especially for large computational grids. Both the creation of the database in the precursor simulation and reading the data from the database in the actual simulation leads to many I/O operations, which can limit the computational efficiency. To alleviate these issues, we have developed a ``concurrent precursor method" in which the precursor and actual simulation are performed simultaneously. Here we use this technique to simulate wind farms without stream-wise periodic boundary conditions while maintaining the advantages of spectral-based numerical methods for high-fidelity LES codes. We first give a general introduction to the LES methodology and the turbine model that is used in section \S \ref{section2} before we discuss the new concurrent precursor method in \S \ref{section3}. In \S \ref{section4} we discuss some sample results, which is followed by a discussion in \S \ref{section5}.

\section{LES model} \label{section2}
In our LES code \cite{cal10,cal11} we consider a neutral ABL and solve the filtered incompressible Navier-Stokes (NS) equations and the continuity equation, i.e.,
\begin{eqnarray}
\partial_{t} \tilde{u}_i + \partial_{j} (\tilde{u}_{i} \tilde{u}_{j}) &=& - \partial_{i} \tilde{p}^{*} - \partial_{j} \tau_{ij}  -\delta_{i1} \partial_{1} p_{\infty}/ \rho + f_i \\
\partial_{i} \tilde{u}_{i} &=& 0
\end{eqnarray}
where $\tilde{u}$ is the filtered velocity field and $\tilde{p}^*$ is the filtered modified pressure equal to $ \tilde{p}/\rho+ \tau_{kk}/3-p_\infty/\rho$. Further, $\tau_{ij}$ is the subgrid-scale stress term. Its deviatoric part ($\tau_{ij} -\delta_{ij} \tau_{kk}/3$) is modeled using an eddy viscosity subgrid-scale model, for which we use the dynamic Lagrangian scale-dependent Smagorinsky model \cite{bou05}. The trace of the SGS stress,  ($\tau_{kk} / 3$) is combined into the modified pressure, as is common practice in LES of incompressible flow. The force $f_i$ is added for modeling the effects of the wind turbines in the momentum equation {\color{black} \cite{jim07,cal10}}. Since simulations are done at very high Reynolds numbers, the bottom surface and the wind-turbine effects are parametrized, while the viscous stresses are neglected {\color{black} \cite{moe84,mas94}}. Our code uses a pseudo spectral discretization, and thus doubly periodic boundary conditions in the horizontal directions. In the vertical direction, the code uses  centered second-order finite differencing. A second-order accurate Adams-Bashforth scheme is used for the time integration. The top boundary uses zero vertical velocity and a zero shear stress boundary condition. At the bottom surface a classic imposed wall stress boundary condition relates the wall stress to the velocity at the first grid point. For this we use the standard log similarity law \cite{moe84,bou05}. The surface roughness $z_{0,lo}=5\times10^{-5} L_z$ with $L_z=2$ km and $\kappa$ is the Von K\'arm\'an constant ($\kappa=0.4$).

The wind turbines are modeled through an actuator disk approach, {\color{black} see also Ref.\ \cite{jim07,cal10,cal11}.} The detailed comparisons with wind-tunnel data presented in Wu and Port\'e-Agel \cite{wu11,wu13} show that except for the near-wake region ({\color{black} up to $3$ to $4D$ behind the turbine}), the actuator (drag)-disk approach yields a good degree of accuracy {\color{black} for the stream-wise velocity in the far wake, i.e. further than $3$ to $4D$ behind the turbine.} The method is based on a drag force ($F_t$) acting in the stream-wise direction ($x_1$) according to
\begin{equation}
F_t = - \frac{1}{2} \rho C_T U^2_\infty \frac{\pi}{4} D^2,
\end{equation}
where $C_T$ is the thrust coefficient, $D$ is the rotor diameter, and $U_\infty$ is an upstream (unperturbed) velocity. This is a good approach when one is modeling a single wind turbine and there are no other interacting bodies in the numerical domain that can make specification of $U_\infty$ ambiguous. When modeling wind farms, it is impossible to define an unperturbed upstream mean velocity since the upstream values are always affected by other upstream wind turbines. It is thus more convenient \cite{mey10} to use the local velocity at the rotor disk $U_d$. It can be related to an equivalent upstream unperturbed velocity through the actuator-disk theory
\begin{equation}
U_\infty = \frac{U_d}{1-a}
\end{equation}
where $a$ is the induction factor. Moreover, modeling the thrust forces acting on the fluid due to its interaction with the rotating blades requires the use of an average disk velocity. It is evaluated from LES by averaging over the disk region, and using a first order relaxation method with a typical time of $10$ seconds, yielding a velocity denoted by $\langle \overline{u}^{T} \rangle_d$ {\color{black} \cite{cal10,mey10}}. Then, the total thrust force can be written as
\begin{equation}
F_t = -\rho \frac{1}{2} C_T^{'} \langle \overline{u}^{T} \rangle_d^2 \frac{\pi}{4}D^2,
\end{equation}
with the subscript $d$ denoting an averaging over the turbine disk region and the superscript $T$ denotes the time averaging. We also define
\begin{equation}
C_T^{'} =\frac{C_T}{(1-a)^2}.
\end{equation}
For the Betz limit (i.e.\ $C_T=8/9$, and $a=1/3$), we obtain $C_T^{'}=2$. In this study we use values which may be found in existing wind turbines \cite{bur01} and prior LES studies \cite{jim07}, i.e.\ we use $C_T=3/4$ and $a=1/4$, which leads to $C_T^{'}=4/3$.

The total thrust force is distributed using an indicator function which is determined during code initialization. First the grid-points that fall within each turbine radius are located and subsequently a Gaussian filter $G(x)=[6 / ( \pi\Delta^2)]^{3/2}\exp(-6 ||x||^2/ \Delta^2)$, where $x$ is the distance from the turbine center and $\Delta^2=h^2 \cdot (\Delta x^2 + \Delta y^2 + \Delta z^2)$ (and $h=1.5$) is used to smooth the indicator function. In order to limit the spatial extent of each turbine, a minimum allowable value of the indicator function is enforced. Subsequently the resulting indicator function is normalized with the volume of the turbine disk in order to make sure that the total applied force is independent of the grid resolution. In each grid-point where the indicator is non-zero a force on the flow that corresponds to the value of the indicator function is applied. This method is similar to the one used in {\color{black} Ref.\ \cite{cal10,cal11}.}

\section{Concurrent precursor method} \label{section3}

\begin{figure}
\centering
\subfigure{\includegraphics[width=0.99\textwidth]{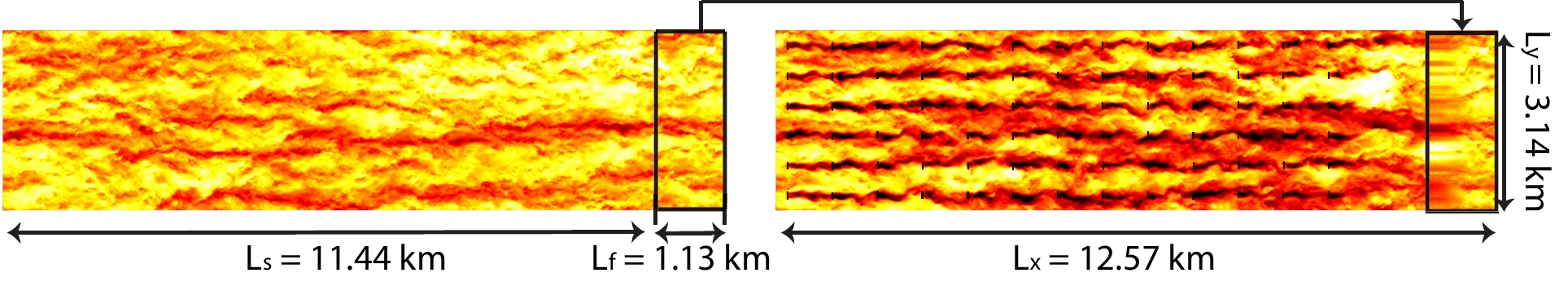}}
    \caption{Instantaneous stream-wise velocity at hub height. Each time step the data at the end of the first domain, in which a turbulent ABL is simulated, are copied to the end of the second domain with the wind turbines where it acts as a turbulent inflow for the first row. See figure \ref{figure2} for a zoom in of the region where both domains meet. The same color scale as in figure \ref{figure2} is used.}
\label{figure1}
\end{figure}

Conceptually the proposed approach is the same as for a precursor simulation {\color{black} \cite{lun98,fer04,tab10}} that is performed before the actual simulation {\color{black} where inflow data is written to disk}. However, now the sampled inflow data from the precursor simulation are not written to file but directly into the memory of the target simulation using MPI (Message Passing Interface) and this limits the required disk space. The direct memory copies also remove the I/O overhead and thereby increase the speed of the {\color{black}simulation}. Since both simulations are executed simultaneously one does not have to wait until the precursor simulation is completed. 

\begin{figure}
\centering
\subfigure{\includegraphics[width=0.6\textwidth]{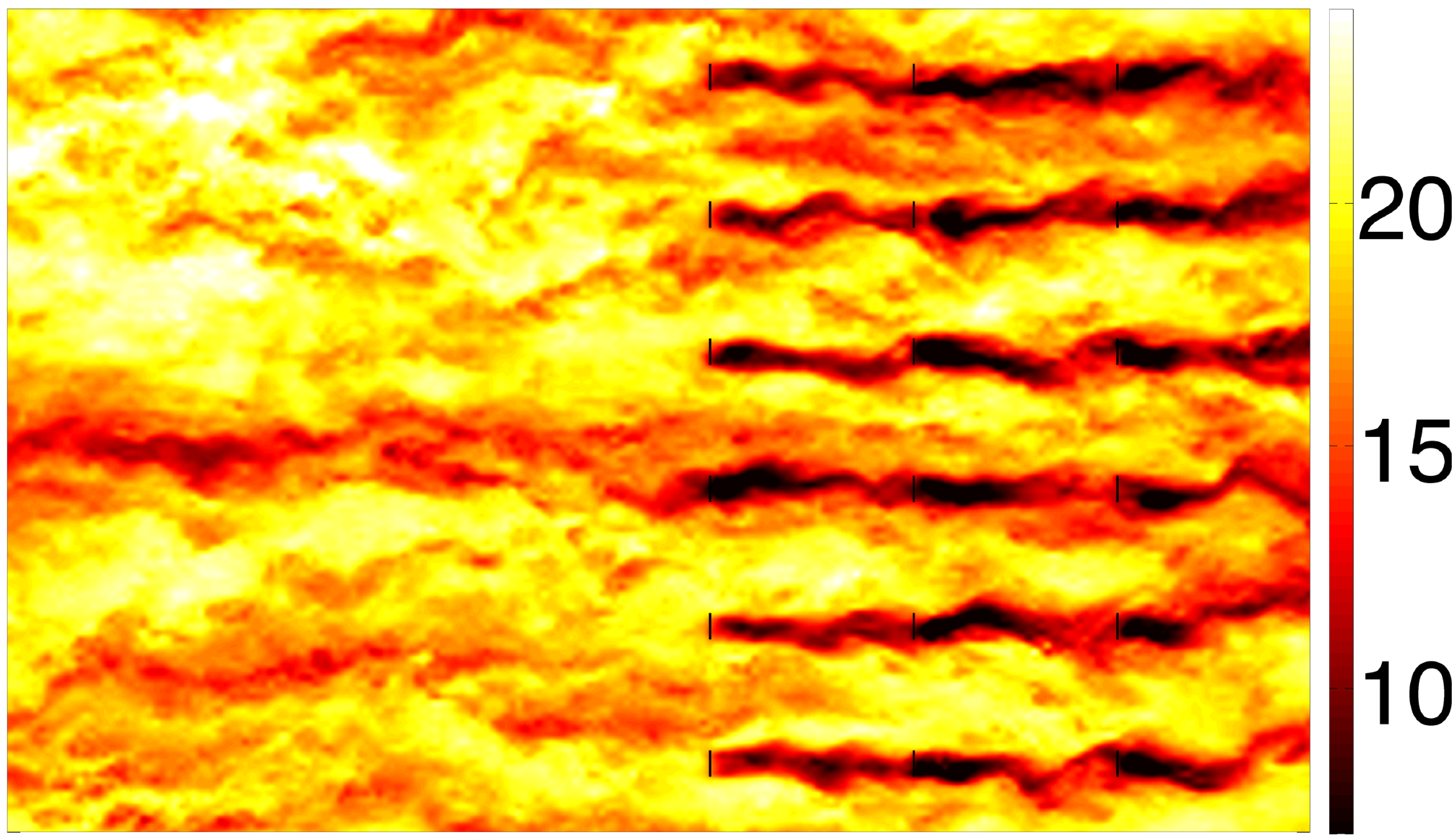}}
    \caption{Zoom in on the region where the two domains meet. Note the smooth transition between the two domains. The color indicates the streamwise velocity in non-dimensional units $u/u_*$.}
\label{figure2}
\end{figure}

\begin{figure}
\centering
\subfigure{\includegraphics[width=0.6\textwidth]{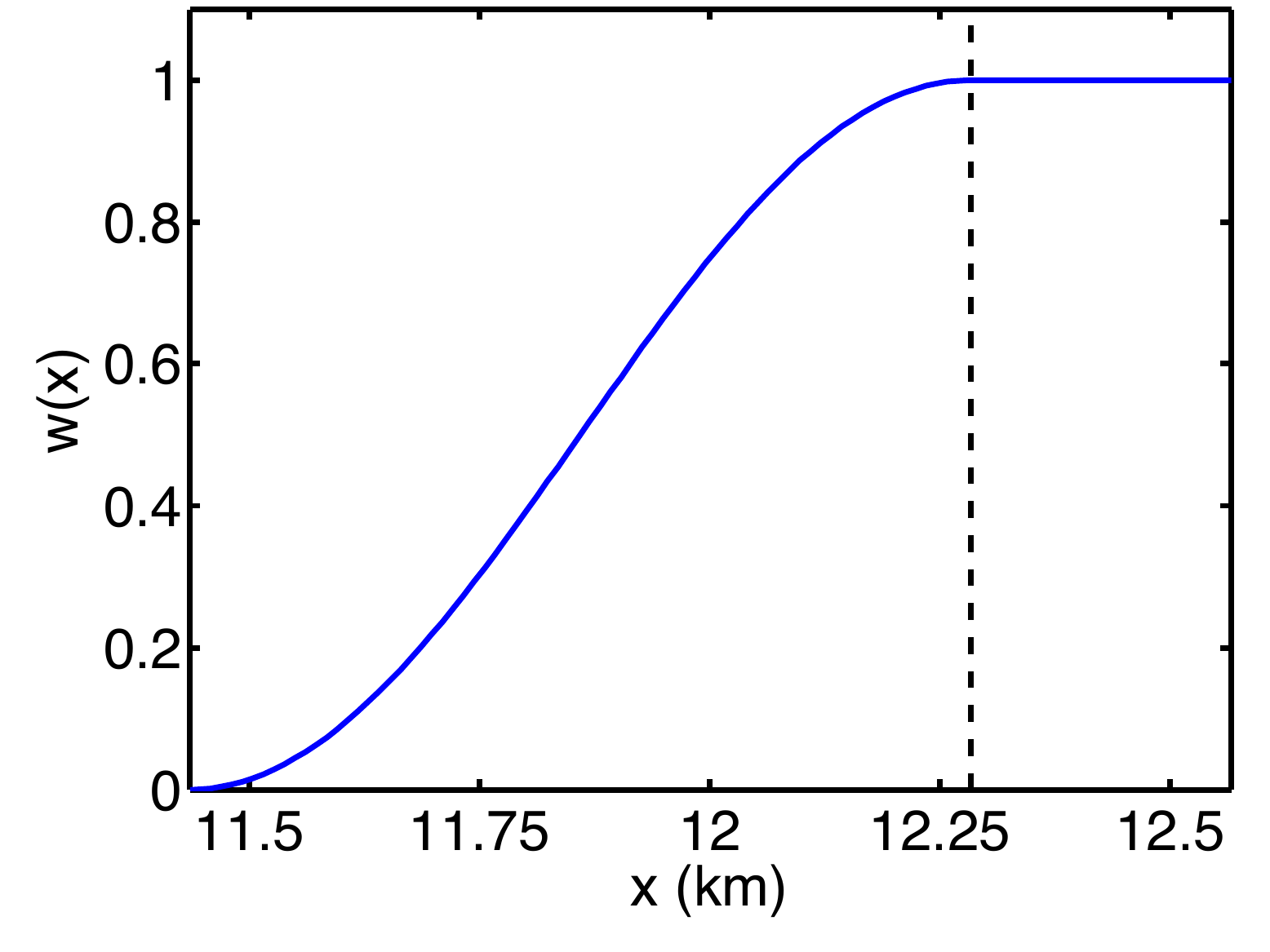}}
    \caption{The weighing function $w$ as function of x. $L_{s} =11.44$ km, $L_{pl} = 12.28$ km (dashed line), and $L_{e} = 12.57$ km.}
\label{figure3}
\end{figure}

In the first domain, a classical turbulent ABL that is driven {\color{black} by an imposed} pressure gradient in the stream-wise direction is considered. In the second domain a finite length wind farm is considered and due to the forced inflow condition there is no need to impose a pressure gradient in this second domain. At the end of each time step a region of length $L_f$ is copied from the first domain to the end of the second domain {\color{black} (from $x=L_s (L_s=L_x-L_f)$ to $x=L_x$)}, see figure \ref{figure1} and \ref{figure2}). {\color{black} Here we used $L_f=0.09L_x$. We note that the best length can depend on the domain length and the case that is considered.} In order to prevent unwanted oscillations in the domain, a fringe region is used towards the outflow of the domain where the wind-turbines are placed \cite{spa88,sch05,che07}. The velocity in the fringe region is determined as
\begin{equation}
{\bf u}_{fr} (x,y,z,t)= w (x) \cdot {\bf u}_{pre} \left(x,y,z,t)+ (1-w(x))\cdot {\bf u}_{LES}(x=L_{s},y,z,t\right),
\end{equation}
where ${\bf{u}}_{pre}$ is the velocity field in the precursor domain's outflow region and ${\bf{u}}_{LES}$ is the velocity in the actual domain. The blending function $w$ indicates the relative importance of the precursor domain in the fringe region. The influence of the precursor domain in the fringe region is gradually increased using the following weighing function
\begin{equation}
w=\frac{1}{2}\left[1 - \cos \left(\pi \frac{x - L_{s}}{L_{pl} - L_{s}}\right) \right] ~~ \mbox{if~~~x} > L_{s} \mbox{~~~and~~~} w= 1 ~~~ \mbox{if~~~x} > L_{pl},
\end{equation}
where $x$ indicates the stream-wise position and $L_{pl}=L_{e}-\frac{1}{4}L_f$, see figure \ref{figure3}. The use of this weighing function in combination with a sufficient length of the fringe region ensures that there is a smooth transition from the velocities that are directly calculated in the domain with the wind turbines to the turbulent inflow velocities obtained from the precursor simulation. Communication between the two domains makes sure that the time advancement in the two domains is synchronized {\color{black} and required to satisfy a CFL number of 0.065 defined as
\begin{equation}
	\mbox{CFL} = \Delta t \mbox{ max } \left(\frac{|u_x|}{\Delta x}, \frac{|u_y|}{\Delta y}, \frac{|u_z|}{\Delta z} \right).
\end{equation}
We also note that the grid spacing of both domains is identical in order to allow the copying of the flow field $\bf{u}$, i.e. the stream-wise, span-wise, and vertical velocity component, without interpolation.}

\section{Sample results}  \label{section4}

\begin{figure}
\centering
\subfigure[]{\includegraphics[width=0.49\textwidth]{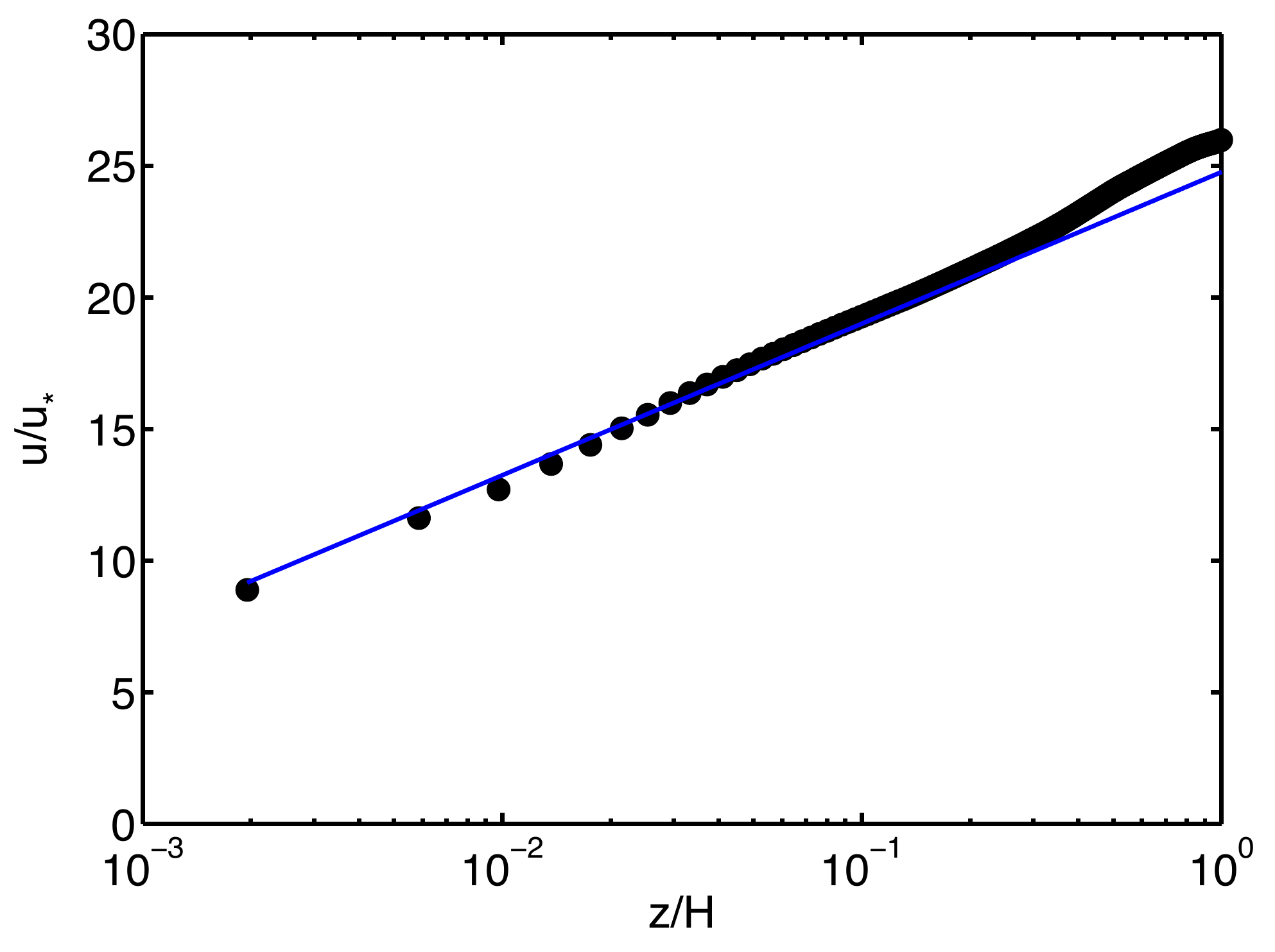}}
\subfigure[]{\includegraphics[width=0.49\textwidth]{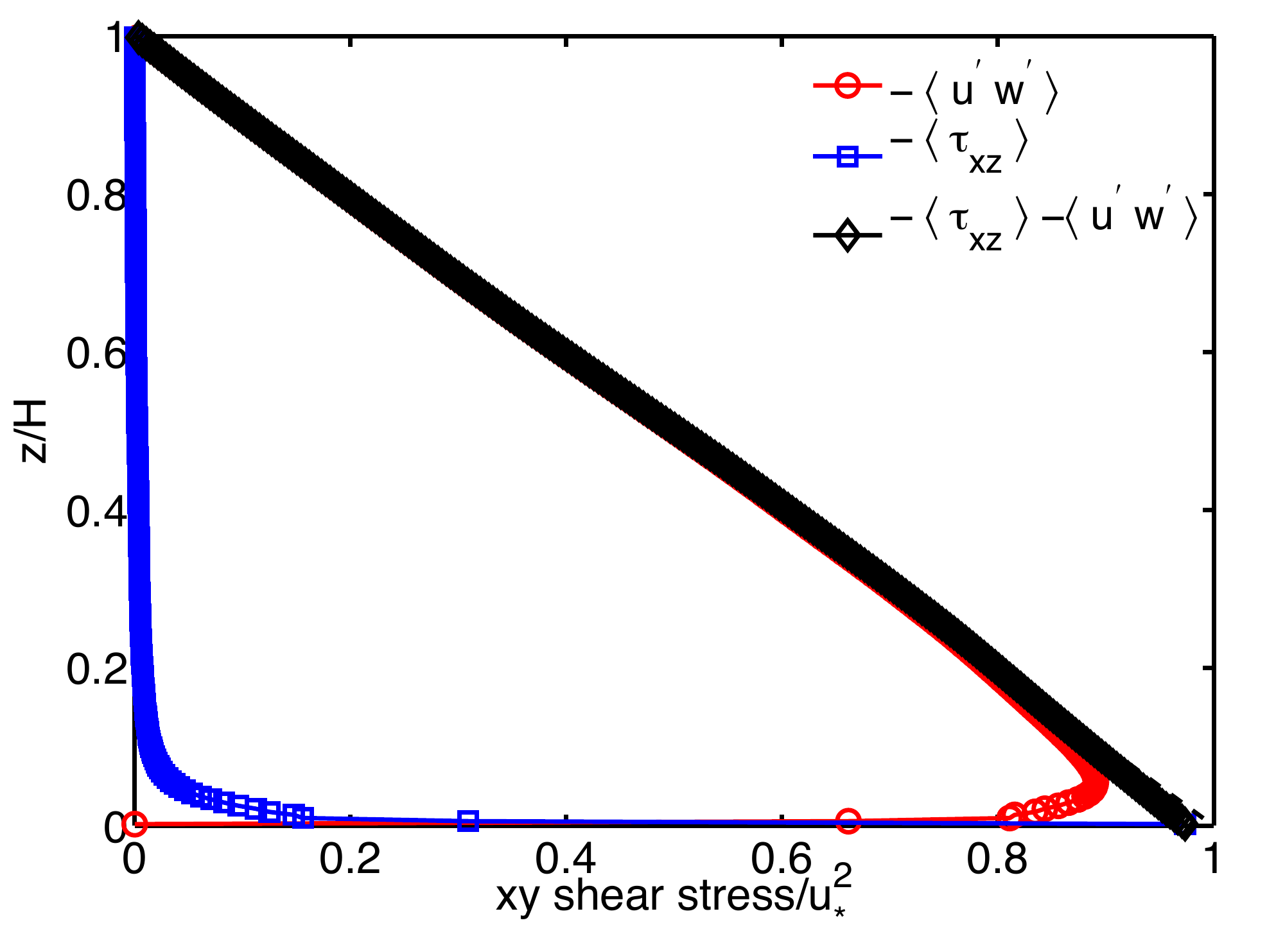}}
    \caption{{\color{black} a) Horizontally averaged stream-wise velocity from the precursor LES (circles) compared with the logarithmic law (line). b) Vertical profiles of the resolved stress ($- \langle u'w' \rangle$, circles) the subgrid scale stress ($-\langle \tau_{xz} \rangle$, squares) and the total stress ($- \langle u'w' \rangle-\langle \tau_{xz} \rangle$, diamonds).}}
\label{figure3b}
\end{figure}

\begin{figure}
\centering
\subfigure{\includegraphics[width=0.6\textwidth]{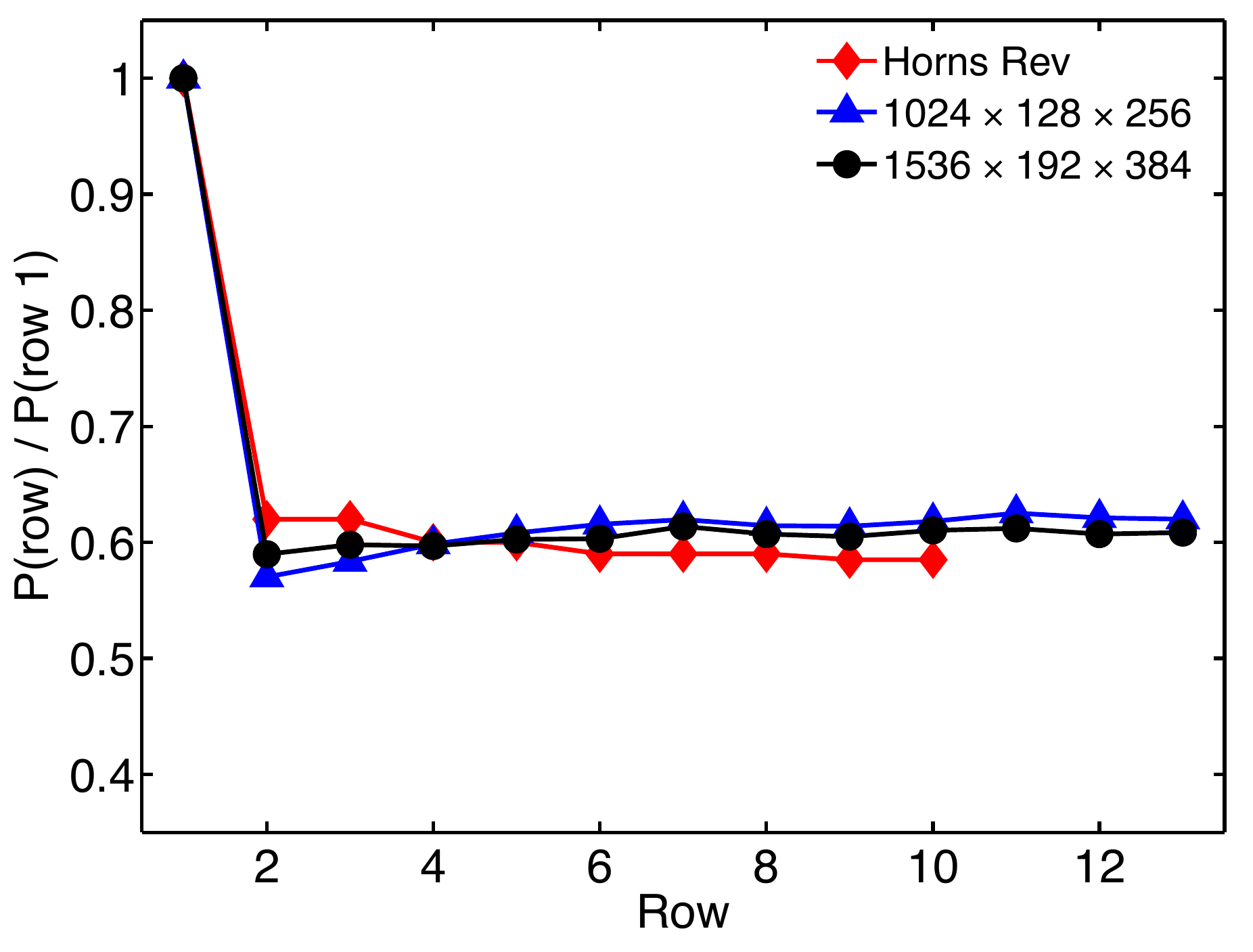}}
    \caption{Power output of subsequent downstream wind turbines compared with the power output of the turbines in the first row. Note that after a strong reduction at the first turbine the power production is nearly constant as function of the downstream position.}
\label{figure4}
\end{figure}

In order to test the concurrent precursor method we performed a simulation of an extended wind farm, which consists of $13$ rows in the stream-wise direction and $6$ turbines in the periodic span-wise direction. The spacing between the turbines is $7.85D$ in the stream-wise direction and $5.24D$ in the span-wise direction. Both domains are $3.14 \times 12.57 \times 2$ km, in the span-wise, stream-wise, and vertical direction, respectively and different resolutions are considered. The length of the fringe region {\color{black} $L_f$} is $1.13$ km in the stream-wise direction and no turbines are placed in the region close to the start of the fringe region. Figure \ref{figure1} reveals the presence of time-dependent streaky flow structures in the precursor domain and therefore this method to generate the inflow condition for the wind farm simulation {\color{black}mimics a closer representation of an ABL} compared to using stored datasets and just sweeping them past, or the use of synthetic turbulence models to generate the inflow conditions. {\color{black} In figure \ref{figure3b} the horizontally averaged stream-wise velocity and the vertical stress profiles obtained from the precursor domain of the simulation performed on the $1024 \times 128 \times 256$ grid, i.e.\ the grid resolution of the precursor domain is thus $512 \times 128 \times 256$. Figure \ref{figure3b}a show that the precursor simulation captures the logarithmic law well, while the vertical stress profiles in figure \ref{figure3b}b reveal that the precursor simulation has reached the statistically stationary state \cite{bou05}.} In figure \ref{figure4} the average power output of the different downstream turbine rows is compared with the average power output of the first row. In agreement with field data the simulation results obtained with this method show for an aligned configuration that there is a strong reduction of the power output of downstream wind-turbines due to wake effects. We also remark that in Ref.\ \cite{ste13} the same field data set has been compared with various other models and simulations of wind farms. Present results shown in Fig.\ \ref{figure4} {\color{black} are in good agreement to} these {\color{black} prior results \cite{san11,bar11,sch12}}. 

\section{Discussion and conclusion} \label{section5}
We have presented a new concurrent precursor technique that can be used for the simulation of finite length wind farms in conjunction with high fidelity spectral codes that are based on periodic boundary conditions. In agreement with field data the simulation results obtained with this method show for an aligned configuration that there is a strong reduction of the power output of downstream wind-turbines due to wake effects. This new method has several advantages with respect to the conventional one in which a precursor simulation is performed before the actual simulation and the turbulent inflow conditions are stored in a database, namely a) from a computational point of view the presented method is beneficial as it prevents the creation of a very large database in which the turbulent inflow data are stored, b) The associated I/O operations, which can limit the computational efficiency of codes, are replaced by direct memory copies that are performed with MPI, and c) From a physical point of view the new method {\color{black} provides a realistic representation of a time-resolved ABL} as it assures that the simulations are not biased due to a fixed and stored inflow condition. In particular, time-evolving streaky structures that are natural in an ABL, but difficult to include in synthetic models or in statically swept spatial fields, are included.\\

{\it Acknowledgements}:  This work is funded in part by the research program 'Fellowships for Young Energy Scientists' (YES!) of the Foundation for Fundamental Research on Matter (FOM) supported by the Netherlands Organization for Scientific Research (NWO), and in part by the US National Science Foundation, grants \#  CBET 1133800 and OISE 1243482. The computations have been performed on our local cluster and the LISA cluster of SARA in the Netherlands.

\end{document}